\documentclass[twocolumn,prl,footinbib,showpacs]{revtex4}

\usepackage{graphicx}
\usepackage{bm}
\usepackage{amsmath}
\usepackage{subfigure}
\usepackage{textcomp}

\begin{document}

\title{Spin and Orbital Rotation of Electrons and Photons via Spin-Orbit Interaction}

\author{C.C. Leary}
\email{cleary@uoregon.edu}
\author{M.G. Raymer}
\author{S.J. van Enk}

\affiliation{Oregon Center for Optics and Department of Physics, University of Oregon, Eugene, OR USA, 97403}

\date{\today}


\begin{abstract}
We show that when an electron or photon propagates in a cylindrically symmetric waveguide, its spin angular momentum (SAM) and its orbital angular momentum (OAM) interact. Remarkably, we find that the dynamics resulting from this spin-orbit interaction are quantitatively described by a single expression applying to both electrons and photons. This leads to the prediction of several novel rotational effects: the spatial or time evolution of either particle's spin/polarization vector is controlled by its OAM quantum number, or conversely, its spatial wavefunction is controlled by its SAM. We show that the common origin of these effects in electrons and photons is a universal geometric phase. We demonstrate how these phenomena can be used to reversibly transfer entanglement between the SAM and OAM degrees of freedom of two-particle states.
\end{abstract}

\maketitle

 It is well known that when an electron propagates in an inhomogeneous potential, its spin angular momentum (SAM) $\hat{\mathbf{S}}$ interacts with the orbital angular momentum (OAM) $\hat{\mathbf{L}}$ associated with its own curvilinear motion. It is also known that when light propagates in a transparent medium with an inhomogeneous refractive index, an analogous effect can take place: its polarization and OAM can interact and alter the propagation characteristics of the light. Several instances of this have been predicted (cf. \cite{Liberman, Onoda, Bliokh_04}), and a few experiments have been done \cite{Dooghin, Hosten, Bliokh_08}. What has not yet been made clear is the extent to which a unified wave-picture description of this spin-orbit interaction (SOI) for both photons (electromagnetic fields) and electrons (matter waves) can be reached.

 In this work we study the dynamics of the SOI from within such a unified framework. Remarkably, we find that the SOI is quantitatively described by a single expression applying to either an electron or a photon propagating in a straight, cylindrically symmetric waveguide geometry. This leads to the prediction of several novel rotational effects for both particle types, in which the particle's spin and orbital degrees of freedom influence one another as it propagates down the waveguide. These phenomena allow for the reversible transfer of entanglement between the SAM and OAM degrees of freedom of two-particle states. To provide deeper insight, we show that the common origin of these effects in electrons and photons is a universal geometric (Berry) phase associated with the interplay between either particle's spin and OAM. This implies that the SOI occurs for any particle with spin, and thereby exists independently of whether or not the particle has mass, charge, or magnetic moment.

 Previous authors have examined the connection between the geometric phase and the SOI for both particle types (cf. \cite{Bliokh_08,Berard_06} and Refs.\ therein). However, the cylindrical geometry we treat here, which supports transversely stationary waves with well-defined OAM that propagate down a \textit{straight} waveguide axis, contrasts with the geometry in nearly all other related studies, which consider a transversely localized beam traveling along either a curved or refracted trajectory (cf. \cite{Onoda, Bliokh_04, Hosten, Bliokh_08, Berard_06, Tomita, Bialynicki-Birula}). Furthermore, these analyses have been limited to the contexts of semi-classical equations of motion in a trajectory (ray) picture, so that a majority of the aforementioned rotational effects, which can be described only via the \textquoteleft{wavefunction\textquoteright} picture, have been missed.

\begin{figure} 
\includegraphics[width=000.48\textwidth]{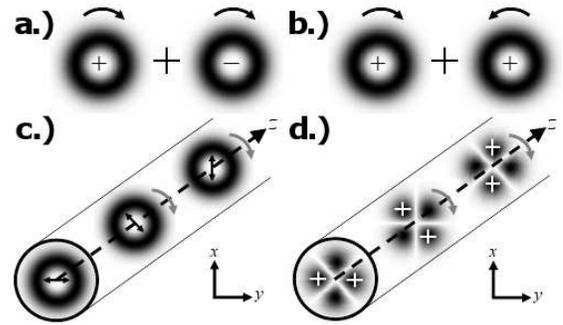}
\vspace{-0.12\textwidth}
\caption{\label{fig:SOI} (a) An OAM eigenstate with $\left|m_{\ell}\right|=2$ in a balanced superposition of $+$ and $-$ SAM states (see equation \eqref{Sup_a}). The $\pm$ signs contained within the transverse spatial profiles indicate the SAM of the contributing state, while the arrows indicate its OAM handedness. (b) A SAM eigenstate in a balanced superposition of right and left-handed OAM states with $\left|m_{\ell}\right|=2$ (see equation \eqref{Sup_b}). When states (a) and (b) propagate down a straight waveguide, the spin (polarization) vector of the state in (a) (see equation \eqref{Evo_a}) and the transverse spatial profile of the state in (b) (see equation \eqref{Evo_b}) exhibit azimuthal rotation, as shown in (c) and (d), with the sense of rotation controlled by the sign of the OAM and SAM quantum numbers, respectively. The straight arrows in (c) denote the orientation of the state's spin (polarization) vector, while the white plus signs in (d) represent relative transverse phase.}
\end{figure}

In order to summarize these SOI phenomena, we introduce the following terminology: we speak of an electron (photon) as being in a SAM eigenstate if its corresponding quantum state (transverse electric field) is an eigenstate of the SAM \textit{z}-component (helicity) operator $\hat{S}_{z}\equiv s\hat{\sigma_z}$ with eigenvalue $s\sigma$. Similarly, an eigenstate of the OAM \textit{z}-component operator $\hat{L}_{z}\equiv-i\partial_{\phi}$ with eigenvalue $m_{\ell}$ is in an OAM eigenstate. Here, $s$ is the particle spin, while $\hat{\sigma_z}$ is the diagonal Pauli matrix, and $\sigma=\pm1$ and $m_\ell=0,\pm1,\pm2,...$ are the SAM and OAM quantum numbers, respectively. 
 
 The first of the aforementioned effects involves an OAM eigenstate in a balanced superposition of SAM states as shown in Fig.\ \ref{fig:SOI}(a). As we will show, when an electron (photon) in this state propagates down the \textit{z} axis of a cylindrical waveguide (see Fig.\ \ref{fig:SOI}(c)), it exhibits a cyclic azimuthal rotation of its spin (polarization) vector in the transverse plane, with the direction of this rotation being controlled by the \textit{sign} of its OAM quantum number $m_{\ell}$, denoted by $\mu\equiv\tfrac{m_{\ell}}{\left|m_{\ell}\right|}=\pm1$. Conversely to this, a second effect involves the directional control of the azimuthal rotation of either particle's \textit{transverse spatial wavefunction} (i.e., its \textquoteleft{orbital\textquoteright} state) by the sign of its SAM quantum number $\sigma$, which occurs when a SAM eigenstate in a balanced superposition of OAM states (see Fig.\ \ref{fig:SOI}(b)) propagates in the waveguide as shown in Fig.\ \ref{fig:SOI}(d). One may think of the former phenomenon as \textit{orbit}-controlled \textit{spin} rotation, and the latter as \textit{spin}-controlled \textit{orbital} rotation. As we discuss later, both of these rotation effects may occur either as a function of space (distance traveled down the guide) or of time, so that combining the above possibilities then yields four distinct, complementary manifestations of the SOI for each particle type.
  
 We note that the spin/polarization rotation of Fig.\ \ref{fig:SOI}(c) is reminiscent of a well-known polarization rotation effect for light first predicted by Rytov \cite{Rytov} and observed by Tomita and Chiao \cite{Tomita}, in whose experiment a single-mode optical fiber was wound into a helical shape so that light propagating in the fiber traced out a closed momentum-space path, thereby giving rise to a polarization rotation induced by a Berry phase. However, we stress that the effect predicted in Fig.\ \ref{fig:SOI}(c) has fundamental differences with the Rytov-Chiao rotation, as the former case involves an OAM eigenstate traveling along the \textit{z} axis, while the latter involves the lowest-order fiber mode (which has \textit{zero} OAM) traveling along a helical trajectory and predicts the \textit{absence} of any polarization rotation if the fiber is made perfectly straight. As we will show, the reason that rotation still occurs for a straight waveguide in the former case is that the OAM states assume the role played by the helical fiber geometry.
    
 The orbital rotation of Fig.\ \ref{fig:SOI}(d) has been previously predicted and observed indirectly for photons in a multi-mode fiber by Zel'dovich et al. \cite{Liberman, Dooghin} (although the aforementioned temporal manifestation was not discussed by them). Aspects of the electron case have been treated previously by us \cite{Leary}. Here we present a single expression, valid for both electrons and photons, which provides a unified description of this orbital rotation effect with the three additional complementary effects discussed above. 
 
  Consider first an electron with mass $m$ and charge $-e$ guided at a non-relativistic velocity by an inhomogeneous but cylindrically symmetric potential $V\left(\rho\right)$ with an effective radius $a$. After normalizing the electron's potential energy $U\left(\rho\right)=-eV\left(\rho\right)$ by its rest energy $mc^{2}$, we may express the resulting quantity as $\tfrac{U\left(\rho\right)}{mc^{2}}=-\big(\tfrac{eV\left(0\right)}{mc^{2}}-\Delta\chi\left(\rho\right)\big)$, where $\Delta=\tfrac{e\left(V\left(0\right)-V\left(a\right)\right)}{mc^{2}}$. In order to ensure transversely bound states, we assume that $\chi\left(\rho\right)$ is zero at the origin and increases monotonically to one at radius $a$, becoming constant thereafter. Due to the \textit{z}-axis translational invariance of the waveguide geometry, a monoenergetic wavefunction in the coordinate basis assumes the traveling-wave form $\Psi\left(\rho,\phi\right)e^{i\left(\beta z-\omega t\right)}$, and the Dirac equation for the electron's wavefunction in the Foldy-Wouthuysen representation assumes the form of a time-independent Schr\"{o}dinger-type equation with a perturbation (c.f. \cite{Greiner}),
\begin{equation} \label{Schrodinger}
\hat{H}_{0}\Psi+\hat{H}'\Psi=\beta^{2}\Psi.
\end{equation}
\noindent Here, the unperturbed operator $\hat{H}_{0}$ has the well-known Schr\"{o}dinger form $\hat{H}_{0}\equiv\nabla_{T}^{2}+k^{2}\left(\rho\right)$, where $\nabla_{T}^{2}$ is the transverse Laplacian, and  $k^{2}\left(\rho\right)=\tfrac{2m}{\hbar^{2}}\left(\hbar\omega+eV\left(\rho\right)\right)$. The solutions of the unperturbed, separable, scalar equation $\hat{H}_{0}\Psi^{\left(0\right)}=\beta_{0}^{2}\Psi^{\left(0\right)}$ are eigenstates of both SAM and OAM, and have the general form
\begin{equation} \label{Eigenstates}
\Psi^{\left(0\right)}=N\psi_{\left|m_{\ell}\right|}\left(\kappa\rho\right)e^{i\left(m_{\ell}\phi+\beta_{0} z-\omega t\right)}\hat{\mathbf{e}}_{\sigma}\equiv\Psi_{\sigma m_{\ell}},
\end{equation}
\noindent where $N$ is a normalization constant, $\kappa$ is the transverse wavenumber, and the spinor $\hat{\mathbf{e}}_{\sigma}$ denotes the spin state of the electron as determined by $\sigma$. 

 The perturbation $\hat{H}'$ in \eqref{Schrodinger} consists of three well-known contributions: a relativistic kinetic energy correction, a spin-orbit interaction correction, and the so-called Darwin term. We are concerned only with the spin-orbit contribution, which for an external electrostatic potential takes the general form $\hat{H}_{SO}=-\tfrac{e}{m c^{2} } \frac{\hat{\mathbf{S}}}{\hbar}\cdot \left(\mathbf{E}\times \frac{\hat{\mathbf{p}}}{\hbar}\right)$, where $\mathbf{E}=-\nabla V\left(\rho\right)$. In \cite{Leary}, we showed that for $\Delta\ll 1$, in first-order perturbation theory $\hat{H}_{SO}$ assumes the form
\begin{equation} \label{Hamiltonian}
\hat{H}_{SO}=-\frac{\Delta}{2}\frac{1}{\rho}\frac{\partial\chi}{\partial\rho}\hat{\sigma}_{z}\hat{L}_{z}
\end{equation}
\noindent under quasi-paraxial conditions.

 Consider now the analogous case of a photon propagating in an inhomogeneous but cylindrically symmetric medium characterized by permittivity $\epsilon\left(\rho\right)$, which can be written in a form similar to that above, $\epsilon\left(\rho\right)=\epsilon\left(0\right)\big(1-\Delta\chi\left(\rho\right)\big)$, where in the photon case $\Delta=\tfrac{\epsilon\left(0\right)-\epsilon\left(a\right)}{\epsilon\left(0\right)}$. In this geometry, the Maxwell equations reduce to a Helmholtz-type wave equation for the transverse electric field, which also assumes the form \eqref{Schrodinger} but with $k^{2}\left(\rho\right)=\omega^{2}\epsilon\left(\rho\right)\mu_{0}$ and $\Psi\to \mathbf{E}_{T}$, where $\mu_{0}$ is the free-space permeability and $\mathbf{E}_{T}$ denotes the transverse electric field. It follows from this that the photonic solutions to the unperturbed equation $\hat{H}_{0}\mathbf{E}_{T}^{\left(0\right)}=\beta_{0}^{2}\mathbf{E}_{T}^{\left(0\right)}$ also have the form \eqref{Eigenstates}, where $\hat{\mathbf{e}}_{\sigma}\equiv\frac{1}{\sqrt{2}}\left(\hat{\mathbf{x}}+ i\sigma\hat{\mathbf{y}}\right)$ (here a vector) denotes the circular polarization (helicity) state of the photon as determined by $\sigma$. The perturbation term in \eqref{Schrodinger} for the photon case has the well-known form \cite{Snyder} $\hat{H}'\mathbf{E}_{T}=\nabla_{T}\left[\nabla_{T}\ln\epsilon\left(\rho\right)\cdot\mathbf{E}_{T} \right]$.
 
 A principal result of this paper is that the perturbation $\hat{H}'$ for the photon case also contains a spin-orbit term, which for $\Delta\ll 1$ (weakly guided light fields) is given by the same expression \eqref{Hamiltonian} as in the electron case, so that the physics of the spin-orbit interaction is completely analogous for electrons and photons in this regime. Specifically, \eqref{Eigenstates} and \eqref{Hamiltonian} allow us to use perturbation theory to calculate the first-order corrections $\delta\beta$ to the propagation constants $\beta_{0}$ due to the SOI for either particle, as
\begin{subequations} \label{dB}
\begin{align}
\delta\beta&=\frac{1}{2\beta_{0}}\left\langle\Psi^{\left(0\right)}\right|\hat{H}_{SO}\left|\Psi^{\left(0\right)}\right\rangle\equiv\frac{1}{2\beta_{0}}\left\langle\mathbf{E}_{T}^{\left(0\right)}\right|\hat{H}_{SO}\left|\mathbf{E}_{T}^{\left(0\right)}\right\rangle 
\nonumber \\
&=-\sigma m_{\ell}\frac{\pi\Delta}{2 \beta_{0}}N^{2}\int{\frac{\partial\chi}{\partial\rho}\left|\psi_{\left|m_{\ell}\right|}\left(\kappa\rho\right)\right|^{2}d\rho} \label{dB_general} \\
&\rightarrow-\sigma \mu\frac{\Delta}{2a}\frac{\left|m_{\ell}\right|}{\beta_{0}a}\left[\pi a^2N^{2}J_{\left|m_{\ell}\right|}\left(\kappa a\right)^{2}\right], \label{dB_step}
\end{align}
\end{subequations}
\noindent where \eqref{dB_step} gives $\delta\beta$ for the special case $\chi\left(\rho\right)=\Theta\left(\rho-a\right)$ with $\Theta$ being the unit step function, and where $J_{\left|m_{\ell}\right|}\left(\kappa a\right)$ is an $m_{\ell}^{\text{th}}$-order Bessel function of the first kind. It follows directly from \eqref{dB} that upon traveling a distance $z$ along the waveguide, a particle in the state $\Psi_{\sigma m_{\ell}}$ picks up a phase of the form $\gamma=\delta\beta z$, which acts as a correction of magnitude $\left|\delta\beta\right|$ to the particle's propagation constant $\beta_{0}$. The sign of this correction is determined by the product $\sigma\mu$, such that the acquiring of the phase $\gamma$ may be expressed via the transformation
\begin{equation} \label{Phase}
\Psi_{\sigma m_{\ell}} \to \Psi_{\sigma m_{\ell}} e^{-i\sigma\mu\left|\delta\beta\right| z}.
\end{equation}
\noindent Deferring the details of our derivation of the Hamiltonian \eqref{Hamiltonian} for the photon case to a future paper, we employ our result \eqref{Phase} to predict the SOI effects shown in Fig.\ \ref{fig:SOI}. We then proceed to apply an intuitive geometric phase-based approach to the special case of a step-profile for $\chi\left(\rho\right)$, which allows us to identify the phase factor in \eqref{Phase} as a universal geometric phase associated with the SOI.

 As discussed previously, we now consider the following distinct balanced superpositions of the monoenergetic, unperturbed eigenstates $\Psi_{\sigma m_{\ell}}$, 
\begin{subequations} \label{Sup}
\begin{align}
&\frac{1}{\sqrt{2}}\Big(\Psi_{\sigma m_{\ell}}+\Psi_{\left(-\sigma\right) m_{\ell}}\Big)\propto e^{im_{\ell}\phi}\left(\hat{\mathbf{e}}_{\sigma}+\hat{\mathbf{e}}_{-\sigma}\right), \label{Sup_a} \\
&\frac{1}{\sqrt{2}}\Big(\Psi_{\sigma m_{\ell}}+\Psi_{\sigma \left(-m_{\ell}\right)}\Big)\propto \cos\left(\left|m_{\ell}\right|\phi\right)\hat{\mathbf{e}}_{\sigma}, \label{Sup_b}
\end{align}
\end{subequations}
\noindent which are represented pictorially in Figs. \ref{fig:SOI}(a) and \ref{fig:SOI}(b). An important similarity between \eqref{Sup_a} and \eqref{Sup_b} is that for each of these superpositions one of the contributing eigenstates has \textit{parallel} SAM and OAM vectors (that is, the product $\sigma m_{\ell}$ is positive), while the other has \textit{anti}-parallel SAM and OAM vectors ($\sigma m_{\ell}$ negative); this allows the $\sigma\mu$-dependent phase accumulation to manifest itself as the particle propagates down the waveguide. According to \eqref{Phase}, upon propagating a longitudinal distance \textit{z}, the respective states \eqref{Sup_a} and \eqref{Sup_b} evolve into 
\begin{subequations} \label{Evo}
\begin{align}
\frac{1}{\sqrt{2}}&\Big(\Psi_{\sigma m_{\ell}} e^{-i\sigma\mu\left|\delta\beta\right| z}+\Psi_{\left(-\sigma\right) m_{\ell}} e^{+i\sigma\mu\left|\delta\beta\right| z}\Big) \nonumber \\
& \propto e^{i m_{\ell}\phi}\left(\hat{\mathbf{e}}_{+}e^{-i\mu\left|\delta\beta\right| z}+\hat{\mathbf{e}}_{-}e^{+i\mu\left|\delta\beta\right| z}\right),
\label{Evo_a} \\
\frac{1}{\sqrt{2}}&\Big(\Psi_{\sigma m_{\ell}} e^{-i\sigma\mu\left|\delta\beta\right| z}+\Psi_{\sigma \left(-m_{\ell}\right)} e^{+i\sigma\mu\left|\delta\beta\right| z}\Big) \nonumber \\
& \propto \cos\left(\left|m_{\ell}\right|\phi-\sigma\left|\delta\beta\right| z\right) \hat{\mathbf{e}}_{\sigma}. \label{Evo_b} 
\end{align}
\end{subequations}
 For a monoenergetic photon, \eqref{Evo_a} describes a linearly polarized OAM eigenstate whose polarization vector rotates with increasing $z$ as shown in Fig.\ \ref{fig:SOI}(c), such that in a Cartesian basis it can be written as $\cos\left(\left|\delta\beta\right| z\right)\hat{\mathbf{x}}+\mu\sin\left(\left|\delta\beta\right| z\right)\hat{\mathbf{y}}$. Similarly, the expectation value of a monoenergetic electron's spin vector, which rotates in an similar manner, is $\langle\hat{\mathbf{S}}\rangle=\tfrac{\hbar}{2}\left[\cos\left(2\left|\delta\beta\right| z\right)\hat{\mathbf{x}}+\mu\sin\left(2\left|\delta\beta\right| z\right)\hat{\mathbf{y}}\right]$.  In contrast to \eqref{Evo_a}, \eqref{Evo_b} describes a SAM eigenstate with a rotating orbital state, which has the same form for both particles (see Fig.\ \ref{fig:SOI}(d)). These effects may be viewed as a spatial beating between two waves of identical frequency $\omega$ but with slightly different propagation constants $\beta_{0}\pm\left|\delta\beta\right|$ which have been split by the SOI. 

\begin{figure} 
\includegraphics[width=000.48\textwidth]{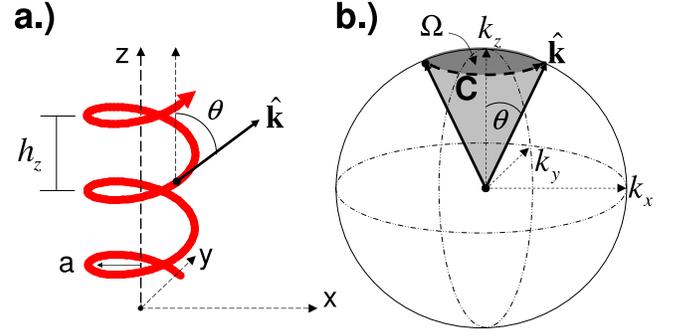}
\vspace{-0.15\textwidth}
\caption{\label{fig:Helix} (a) Helical particle trajectory with constant radius $a$ and pitch $h_{z}=\tfrac{2 \pi a}{\tan\left(\theta\right)}$.
(b) The curve $C$ traced out by $\hat{\mathbf{k}}$ on the momentum-space unit sphere subtends solid angle $\Omega$.}
\end{figure}
 
 As we have mentioned, each rotational SOI manifestation may also occur as a function of time \footnote{For related free-space effects, see S. J. van Enk and G. Nienhuis, Phys. Rev. A \textbf{76}, 053825 (2007).}. In this case, the spin and orbital rotation effects may be viewed as \textit{temporal} beating between waves with identical propagation constants $\beta_{0}$ but different \textit{frequencies} $\omega_{0}\pm\left|\delta\omega\right|$, so that \eqref{Evo} still holds provided that the quantity $\left|\delta\beta\right|z$ is everywhere replaced by $\left|\delta\omega\right|t$. 
 
 Proceeding to the geometric phase approach, we now consider an electron or photon propagating along a well-defined helical trajectory of constant radius $a$ and helix pitch $h_{z}=\tfrac{2 \pi a}{\tan\left(\theta\right)}$ as shown in Fig.\ \ref{fig:Helix}(a), where $\theta$ is the angle that the unit momentum vector $\hat{\mathbf{k}}$ makes with the rotation axis of the helix. In \cite{Bialynicki-Birula}, Bia\l ynicki-Birula and Bia\l ynicka-Birula studied an arbitrary particle with spin traveling along such a classical trajectory, in which the adiabatically varying vector $\hat{\mathbf{k}}$ traces out a closed circular curve $C$ on the momentum space unit sphere, as shown in Fig.\ \ref{fig:Helix}(b).  They found that the particle accumulates a geometric phase associated with the parallel transport of its spin due to its curvilinear motion, which after a single helical round trip can be written as $\gamma=-\lambda\mu_{\text{h}}s\Omega$. Here, $\lambda=\pm1$ denotes the handedness of the particle helicity, $\mu_{\text{h}}=\pm1$ accounts for the handedness of the helical trajectory, $s$ is the particle spin as already defined, and $\Omega$ is the solid angle subtended by the circle $C$ as seen from the origin, which may be expressed as $\Omega=4\pi\sin^{2}\left(\tfrac{\theta}{2}\right)$ (see Fig.\ \ref{fig:Helix}(b)). In order to extend this result to an arbitrary number of helical cycles, we multiply $\gamma$ by $\tfrac{z}{h_{z}}$, so that $\gamma$ may be written as a function of the axial particle position $z$: $\gamma=-\lambda\mu_{\text{h}}s\frac{\Omega}{h_{z}}z$. 
 
 We now apply this result to a photon (electron) traveling in a helical path near the interface of a step-index (potential) waveguide with normalized step height $\Delta\ll1$, such that the pitch angle $\theta\ll 1$ approaches the critical angle for total internal reflection. In this case, we have $\theta\approx\sqrt{\Delta}$ (cf. \cite{Snyder}) so that $\Omega\approx\pi\Delta$ and $h_{z}\approx \frac{2\pi a}{\theta}$, which yields $\gamma\approx-\lambda\mu_{\text{h}}s\frac{\Delta}{2a}\theta z$. Furthermore, in this paraxial regime we may associate the helicity handedness with the spin direction along the \textit{z} axis, and the helix handedness with the sign of the OAM quantum number, so that $\lambda\to\sigma$ and $\mu_{\text{h}}\to\mu$, respectively. With these replacements, the accumulated geometric phase $\gamma$ then gives rise to the effective propagation constant shift $\delta\beta_{\text{geo}}=\frac{\gamma}{z}$,
\begin{equation} \label{GP_Parax}
\delta\beta_{\text{geo}}=-\sigma\mu s\frac{\Delta}{2a}\theta.
\end{equation}
Although the geometric phase result \eqref{GP_Parax} does not coincide exactly with the more accurate perturbative result \eqref{dB_step}, their forms are seen to be strikingly similar when it is recognized that the factor in square brackets in \eqref{dB_step} is of order one when $\left|m_{\ell}\right|$ is near its maximal allowed value. Furthermore, in the classical limit of large OAM ($\left|m_{\ell}\right|\gg 1$), corresponding to the above helical path in a large waveguide, $\pi a^2N^{2}J_{\left|m_{\ell}\right|}\left(\kappa a\right)^{2}\to 1$ and $\frac{\left|m_{\ell}\right|}{\beta a}\to\theta$ in \eqref{dB_step}. Equations \eqref{GP_Parax} and \eqref{dB_step} are therefore in complete agreement for photons, where $s=1$. For electrons however, $s=\frac{1}{2}$, for which \eqref{GP_Parax} yields half of the shift given by the more rigorous perturbative calculation. This apparent discrepancy can be explained by recalling that the result \eqref{GP_Parax} \textit{assumed} parallel transport of the spin for both particles. Although this assumption is consistent with our result for photons, our SOI Hamiltonian \eqref{Hamiltonian} actually causes the electron spin vector $\langle\hat{\mathbf{S}}\rangle$ to precess at \textit{twice} the rate of the photon polarization vector (cf. \eqref{Evo_a} and subsequent discussion). One therefore expects the accumulated phase due to the electron spin evolution to be precisely double the amount predicted by \eqref{GP_Parax}, as we find. We conclude that the spin-orbit interaction dynamics of the electron and photon are identical to first order in perturbation theory and have a common geometric origin, with the role of the electron's potential energy $U\left(\rho\right)$ being played by the permittivity $\epsilon\left(\rho\right)$ in the photon case. 

The effects in Figs.~\ref{fig:SOI}(c) and \ref{fig:SOI}(d) allow for the reversible transfer of entanglement between the SAM and OAM degrees of freedom of two-particle states. Denoting the single-particle state in \eqref{Sup_b} as $\left|\sigma\; HG_{11}\right\rangle$ ($\left|m_{\ell}\right|=2$), we consider a purely polarization-entangled Bell state with two photons in spatially separated $HG_{11}$ (Hermite-Gauss-like) spatial modes, $\left|+\; HG_{11}\right\rangle\left|-\; HG_{11}\right\rangle-\left|-\; HG_{11}\right\rangle\left|+\; HG_{11}\right\rangle$. According to \eqref{Evo_b}, for $\left|\delta\beta\right| z=22.5$\textdegree $ $ this state will evolve under the SOI (in separate waveguides) to $\left|+\; HG^{+}_{11}\right\rangle\left|-\; HG^{-}_{11}\right\rangle-\left|-\; HG^{-}_{11}\right\rangle\left|+\; HG^{+}_{11}\right\rangle$, where the single-photon state $\left|\sigma\; HG^{\pm}_{11}\right\rangle$ denotes a photon whose spatial state has been rotated $\pm 22.5$\textdegree $ $ from $\left|\sigma\; HG_{11}\right\rangle$. By employing wave plates and spatial mode converters, it is possible to transform this state into $\left|D\; LG_{+2}\right\rangle\left|A\; LG_{-2}\right\rangle-\left|A\; LG_{-2}\right\rangle\left|D\; LG_{+2}\right\rangle$, where $D$ and $A$ stand for \textquoteleft{diagonal\textquoteright} and \textquoteleft{anti-diagonal\textquoteright} (oriented at $\pm 45$\textdegree $ $) linear polarization, and $LG_{\pm 2}$ stands for an OAM (Laguerre-Gauss-like) eigenstate. Finally, a second SOI interaction described by \eqref{Evo_a} evolves the state into $\left|H\; LG_{+2}\right\rangle\left|H\; LG_{-2}\right\rangle-\left|H\; LG_{-2}\right\rangle\left|H\; LG_{+2}\right\rangle$, where $H$ stands for horizontal polarization. This is a purely OAM-entangled Bell state. 

This work was supported by NSF grant PHY-0554842.

\end{document}